\documentclass[12pt]{iopart}

\usepackage{hyperref}
\usepackage{cleveref}
\usepackage{graphicx}
\usepackage{booktabs}
\usepackage{float}
\usepackage[raster]{tcolorbox}
\usepackage[export]{adjustbox}
\usepackage{caption}
\usepackage{subcaption}

\hypersetup{
    colorlinks=true,
    linkcolor=black,
    urlcolor=cyan,
    citecolor=blue
}

\usepackage[sorting=none]{biblatex}
\begin{document}
\newcommand{\red}[1]{\textcolor{red}{#1}}
\tcbset{}

\title[A graduate laboratory experiment to set up a photon-counting detector using MKIDs]{A graduate laboratory experiment to set up a photon-counting detector using MKIDs}

\author{Pietro Campana, Rodolfo Carobene, Eleonora Cipelli, \\Marco Gobbo, Aurora Perego, Davide Vertemati \footnote{All authors have contributed equally to this paper.}}

\address{Physics Department, University of Milano-Bicocca, Milano, Italy.}

\ead{p.campana1@campus.unimib.it}

\vspace{10pt}

\begin{abstract}

This paper presents a new laboratory activity aimed at developing knowledge and expertise in microwave applications at cryogenic temperatures. The experience focuses on the detection of infrared photons through Microwave Kinetic Inductance Detectors (MKIDs). The experimental setup, theoretical concepts, and activities involved are detailed, highlighting the skills and knowledge gained through the experience.
This experiment is designed for graduate students in the field of quantum technologies.

\end{abstract}

\section{Introduction}
Over the past two decades, there has been a significant increase in research and development of new detectors that operate at cryogenic temperatures leveraging the phenomenon of superconductivity~\cite{Booth1996, Morozov2021}. While detectors based on semiconductors are an established technology that has already achieved remarkable performance~\cite{Hewitt1994}, the significantly smaller band gap of superconductors allows for higher sensitivities and better energy resolutions. Recent and promising devices of this kind are Microwave Kinetic Inductance Detectors (MKIDs)~\cite{Ulbricht2021}, first proposed in 2003 by researchers from the California Institute of Technology and the Jet Propulsion Laboratory~\cite{Day2003}. An MKID consists of a superconducting resonator whose resonant frequency changes in response to the absorption of photons. In addition to their high sensitivity when used as photon detectors, MKIDs can be arranged in multiplexed arrays. This allows for the readout of multiple detectors using a single electrical connection carrying signals at different frequencies. In a cryogenic experiment, multiplexing can be used to simplify the experimental setup inside the cryostat, reducing the heat carried by electrical connections and thus addressing one of the major limitations in constructing large arrays of cryogenic devices. Among the various superconducting detectors employed in the infrared to ultraviolet range, MKIDs stand out as a noteworthy solution thanks to their ease of fabrication, simple multiplexing scheme, and fast response times, having already been employed in arrays of $\sim 10$k pixels for astrophysics research~\cite{Ulbricht2021}. Furthermore, their simple design and control make them a valuable didactic tool for researchers and students looking to familiarize themselves with many fundamental techniques and laboratory instruments in the superconducting quantum technologies field.

This paper introduces a novel course conducted at the Cryogenics Laboratory of the University of Milano-Bicocca and its experimental realization. The course is designed to offer graduate students an overview of cryogenics, microwave electronics, and low-temperature detectors, while also serving as an introduction to the real research environment, where students are allowed to find solutions on their own. As a result, this article represents the outcome of the students' work and not a guide that was provided to them at the start of the experience.
Leveraging the resources of a previous work~\cite{mezzena_development_2020}, a photon detector employing MKIDs was set up and characterized over the course of two semesters.

While some of the instruments may be missing in the laboratories of many institutions, it is worth mentioning that with the advent of commercial RFSoC~\cite{rfsoc4x2,Baldwin2022,Bracken2022} digital boards, there exist low-cost alternative solutions at least for the readout systems. Moreover, while the experience described in this article employs a custom MKIDs chip, it would be possible to implement a similar experiment (at least partially) with other types of superconducting resonators~\cite{McRae2020}.

\begin{figure}[H]
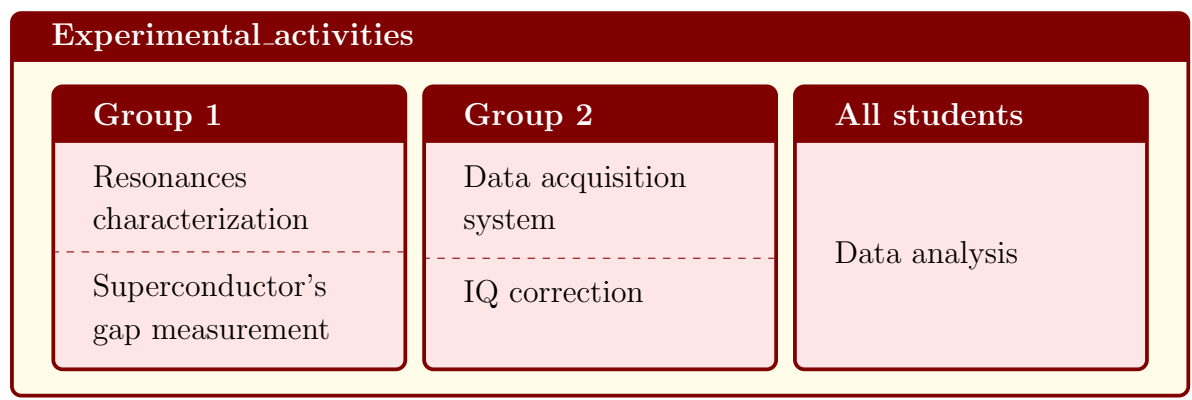

    \begin{tcboxeditemize}[raster columns=3,
        raster equal height,
        colframe=red!50!black,
        colback=red!10!white,
        colbacktitle=red!50!black,
        fonttitle=\bfseries,
        valign = center, 
        space =0.5 ]
        {colback=yellow!10, colframe=red!50!black, fonttitle=\bfseries,title=Experimental\_activities}
\tcbitem[adjusted title = Group 1, valign = center, space =0.5 ] 
Resonances \\ characterization
\tcblower Superconductor's gap measurement
\tcbitem[adjusted title = Group 2]

Data acquisition\\ system
\tcblower  IQ correction \\
\tcbitem[adjusted title = All students] 
Data analysis
    
\end{tcboxeditemize}
\captionsetup{name=Box}
\caption{Outline of the experimental activities.}
\label{modules_chart}
\end{figure}

As shown in box \ref{modules_chart}, the experiment was organized into different activities that could be carried out separately. The students were divided into two groups, with the first one focusing on the characterization of the MKID's resonances and superconductor band gap, while the second group worked on the data acquisition system and the calibration of the raw data acquired with the custom circuit used in photon detection. Finally, data analysis carried out by both groups together confirmed the detection of a small mean number of photons.

The article is structured as follows: in \cref{sec:theory}, the theory necessary to perform the experiment is presented, while \cref{sec:setup} describes the experimental setup. Finally, \cref{sec:activities} details all the activities and the achieved results.

\setcounter{figure}{0}

\section{Theoretical background}\label{sec:theory}
The superconducting properties of a material arise from the formation of Cooper pairs due to interactions between electrons and phonons. The energy required to break a Cooper pair is $2\Delta$, where $\Delta$ is the superconductor's energy gap, which is related to the critical temperature $T_c$~\cite{Bardeen1957, Mattis1958}:
\begin{equation}\label{eq:delta_tc}
    2\Delta \approx 3.5 k_B T_c
\end{equation}
When Cooper pairs are broken, the resulting electrons are called quasiparticles.
In a metal, the mobility of quasiparticles is limited by their short scattering distance. On the other hand, Cooper pairs can move freely and possess an inertia that causes a delay in their response to a changing field, giving rise to a kinetic inductance term $L_k$ in addition to the conventional magnetic inductance $L_m$. 
Kinetic inductance can be exploited to build cryogenics detectors.
In particular, Microwave Kinetic Inductance Detectors (MKIDs) are LC resonators whose resonant frequency is sensitive to the interaction with photons. 

To study the response of MKIDs, we must first consider them as components in a Radio-Frequency (RF) network.
An RF network can be described in terms of the relations between the incident and reflected waves at each of its ports, which can be organized in a complex-valued matrix known as the S-matrix. 
The response of a two-port device can be expressed through the forward gain parameter $S_{21}=\frac{V_2^-}{V_1^+}$, where $V_2^-$ is the wave transmitted across it and $V_1^+$ is the incident one~\cite{pozar2011microwave}.

MKIDs are usually capacitively coupled to a transmission line to measure their response. 
For an ideal capacitively coupled resonator of resonant frequency $f_0$, $S_{21}$ is given by~\cite{Schlaerth}:

\begin{equation}\label{eq:s21_ideal}
    S_{21}(f)=1-\frac{Q_{tot}}{Q_c}\frac{1}{1+2jQ_{tot}\frac{f-f_0}{f_0}}
\end{equation}
where $Q_{tot}$ is the resonance's total quality factor, which represents the energy storage efficiency of the resonator, determining the sensitivity of the device. It is defined as the ratio of the stored energy to the energy dissipated per cycle, or equivalently as the ratio of the resonant frequency to the resonance's bandwidth. The quality factor has two contributions:
\begin{equation}\label{eq:q_relations}
    \frac{1}{Q_{tot}}=\frac{1}{Q_i}+\frac{1}{Q_c}
\end{equation}
where $Q_c$ accounts for the capacitive coupling to the feedline, while $Q_i$ accounts for all other losses intrinsic to the resonator, including the resistive contribution due to quasiparticles and the presence of two-level systems on the material's surface~\cite{GaoTLS}.

Measuring the internal quality factor of an MKID at different temperatures provides a way to determine its $T_c$ from \cref{eq:delta_tc} by indirectly measuring $\Delta$. 
Given $\Delta_0$ the energy gap at $T=0$ K, at low frequencies $\hbar \omega \ll \Delta_0$ and low temperatures $k_B T \ll \Delta_0$, the complex conductivity $\sigma=\sigma_1 + j\sigma_2$ of a superconductor can be approximated by analytical formulas:

\begin{equation}\label{eq:sigma1}
    \frac{\sigma_1 (\omega)}{\sigma_n} \approx \frac{4\Delta}{\hbar \omega} e^{-\frac{\Delta_0}{k_B T}} \sinh (\xi) K_0 (\xi), \qquad \xi=\frac{\hbar \omega}{2k_B T},
\end{equation}
\begin{equation}\label{eq:sigma2}
    \frac{\sigma_2 (\omega)}{\sigma_n} \approx \frac{\pi\Delta}{\hbar \omega}  \left[ 1 - 2e^{-\frac{\Delta_0}{k_B T}} e^{-\xi} I_0 (-\xi) \right]
\end{equation}
where $I_0$ and $K_0$ are the modified Bessel functions of the first and second kind, while $\sigma_n$ is the normal (non-superconductive) conductivity of the metal.
The complex conductivity determines the temperature dependence of the internal quality factor:
\begin{equation}\label{eq:invqi_for_gapfit}
    \frac{1}{Q_i(T)} = \frac{1}{Q_i(0)}+\frac{\alpha \sigma_1(T,\Delta)}{2\sigma_2(T, \Delta)},  \qquad  \alpha = \frac{L_k}{L_k + L_m}
\end{equation}
Therefore, the energy gap can be estimated by substituting $\sigma_1$ and $\sigma_2$ with their approximate expressions (\cref{eq:sigma1,eq:sigma2}) and fitting a set of $Q_i$ values measured at different temperatures in the appropriate range. On the other hand, the kinetic inductance fraction $\alpha$ can be computed with simulations.

The thermal state of a superconductor is characterized by equilibrium populations of Cooper pairs, quasiparticles, and phonons. The interaction with a photon of energy greater than $2\Delta$ initiates a four-stage energy downconversion process ~\cite{Morozov2021, Kozorezov2000}. Initially, few energetic photoelectrons are generated, mostly through electron-electron scattering. As they thermalize, electron-phonon interactions become dominant, increasing the population of quasiparticles through Cooper pair breaking mediated by energetic phonons. The quasiparticle population subsequently concentrates closer to the gap energy $\Delta$. Finally, only phonons with $E<2\Delta$ are present, and the system returns to thermal equilibrium through phonon-phonon scattering, quasiparticle recombination, and phonon escape. In the resonator of an MKID, this process results in a number of Cooper pairs proportional to the energy of the absorbed photon being temporarily broken into quasiparticles. This leads to an increase in the kinetic inductance, which alters the resonant frequency of the MKID before returning to thermal equilibrium.
\begin{figure}
    \centering
    \subfloat[\centering \texttt{}]{{\includegraphics[height=6.5cm]{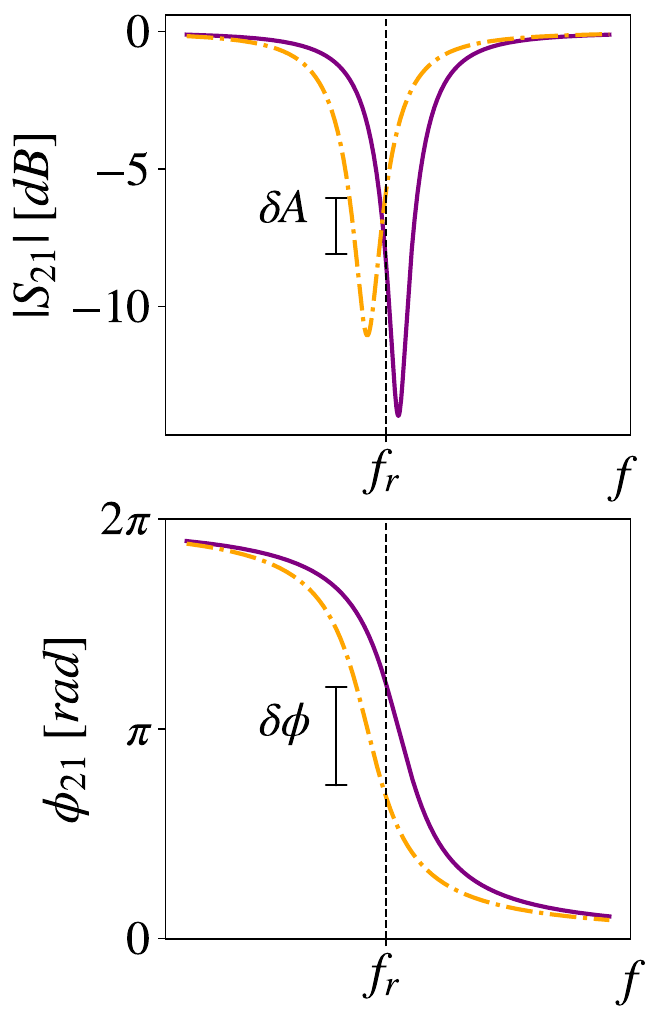}}}\hfill
    \subfloat[\centering\texttt{}]{
    {\includegraphics[height=6.5cm]{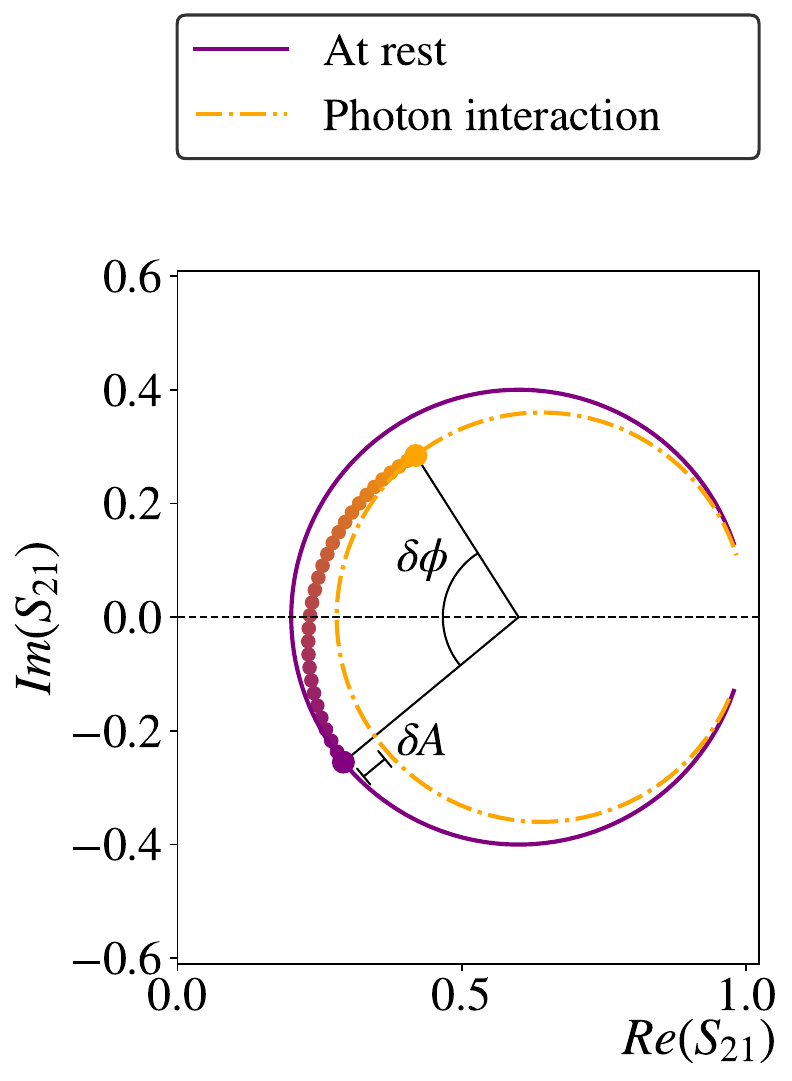}}}\hfill
    \subfloat[\centering\texttt{}]{{\includegraphics[height=6.5cm]{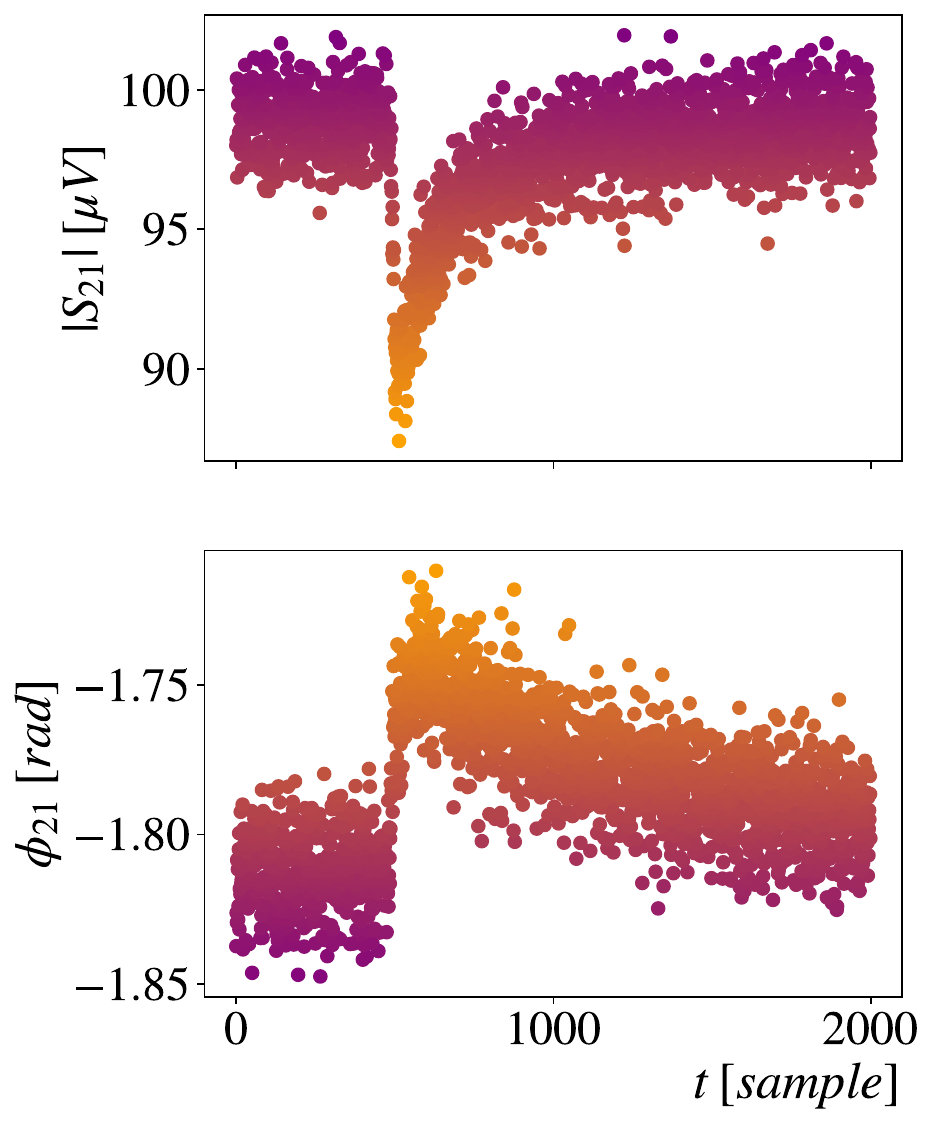}}}
    \caption{Working principle of an MKID. (a) The resonance profile in the frequency domain (purple line). The interaction with a photon alters its frequency and shape (dashed orange line).  If the MKID is probed at a fixed frequency $f_r$, a shift both in amplitude $\delta A$ and in phase $\delta \phi$ can be observed. (b) In the complex plane, the shift appears as moving from a point on the outer (purple) circle to a point on the smaller one (dashed orange line), tracing the path followed by the plotted dots. (c) Example of a raw signal's amplitude and phase in the time domain. The data points are color-coded according to the number of quasiparticles excited by the interaction with a photon, determining the position on the path traced between the two limiting cases depicted in (a) and (b).}
    \label{fig:MKID_shift}
\end{figure}

\Cref{fig:MKID_shift} illustrates how the energy deposited by a photon interacting with an MKID affects its response when considering $S_{21}$~\cite{Gao2008}. To detect the shift of the resonance, the MKID is probed at a fixed frequency while measuring the phase and amplitude of $S_{21}$. In the complex plane, where resonances correspond to circles, the interaction with a photon generates a trajectory between a point on an outer circle and a point on a smaller one and back.
The resulting observed amplitude shift $\delta A$ is proportional to the energy deposited in the resonator. However, the phase excursion $\delta \phi$ typically generates a much larger and more identifiable signal than the amplitude, even if it presents excess noise due to the influence of the two-level systems acting only tangentially to the circles. Since an estimator proportional to the energy can also be easily obtained from $\delta \phi$, the readout of MKIDs often relies on the measurement of the phase of $S_{21}$.
Moreover,  the increased number of quasiparticles generated in the interaction dissipates more energy, reducing the internal quality factor.
In addition to the shift of the resonance's frequency, this also results in a slight change in the shape of its profile.

\section{Experimental setup}\label{sec:setup}

\begin{figure}
    \centering
    \subfloat[\centering \texttt{}]{{\includegraphics[height=6.5cm]{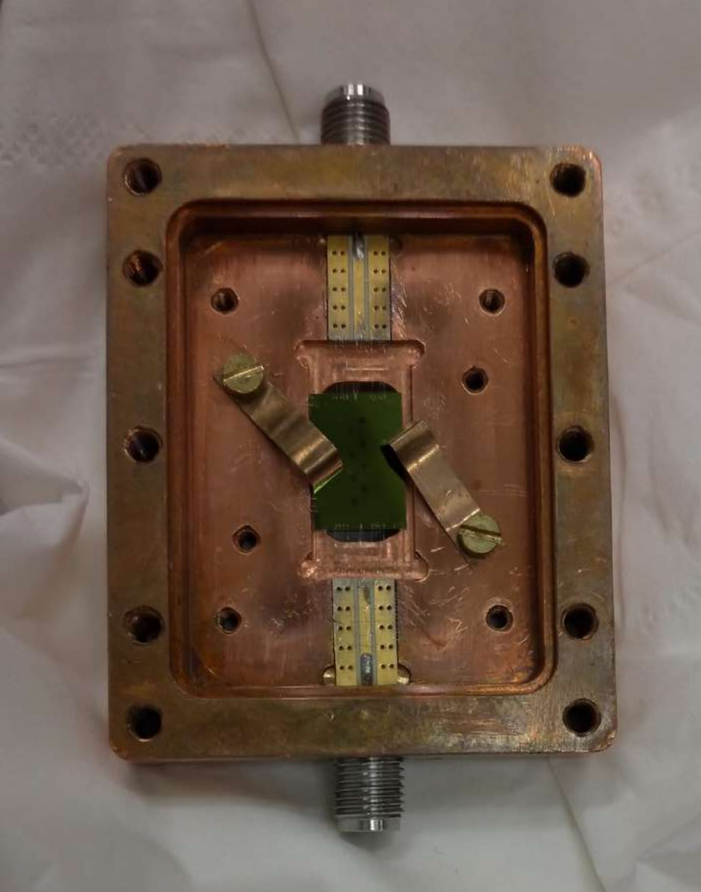}}} \hspace{2cm}
    \subfloat[\centering\texttt{}]{{\includegraphics[height=6.5cm]{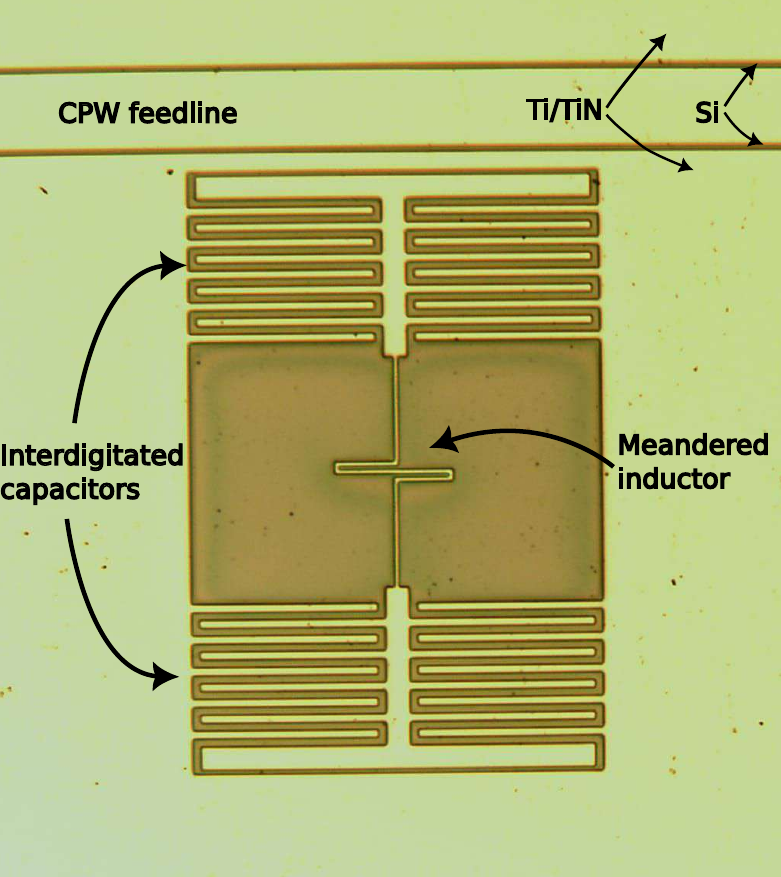}}}
    \caption{(a) The MKIDs chip in the copper holder needed to attach it to the mixing chamber plate. It contains eight resonators placed in a grid. (b) Zoom-in view of a single resonator.}
    \label{mkids_pic}
\end{figure}

The detector consists of a custom chip containing 8 MKIDs with different resonant frequencies. The resonators (see \cref{mkids_pic}) were fabricated using interdigitated capacitors~\cite{Kapoor2020} and are coupled to a shared Coplanar Waveguide (CPW).
The chip, manufactured by Fondazione Bruno Kessler (FBK), is made of a silicon substrate on which alternating layers of titanium and titanium nitride (TiN) are deposited~\cite{Leduc2010}.
Since MKIDs are superconducting detectors, they operate at temperatures significantly below the critical point of TiN.
To achieve this, the chip is placed inside a copper holder and attached to the mixing chamber plate of a dilution cryostat manufactured by Oxford Instruments, which is capable of reaching a minimum temperature of $\sim 40$ mK.

\begin{figure}
    \centering
    \includegraphics[width=0.85\textwidth]{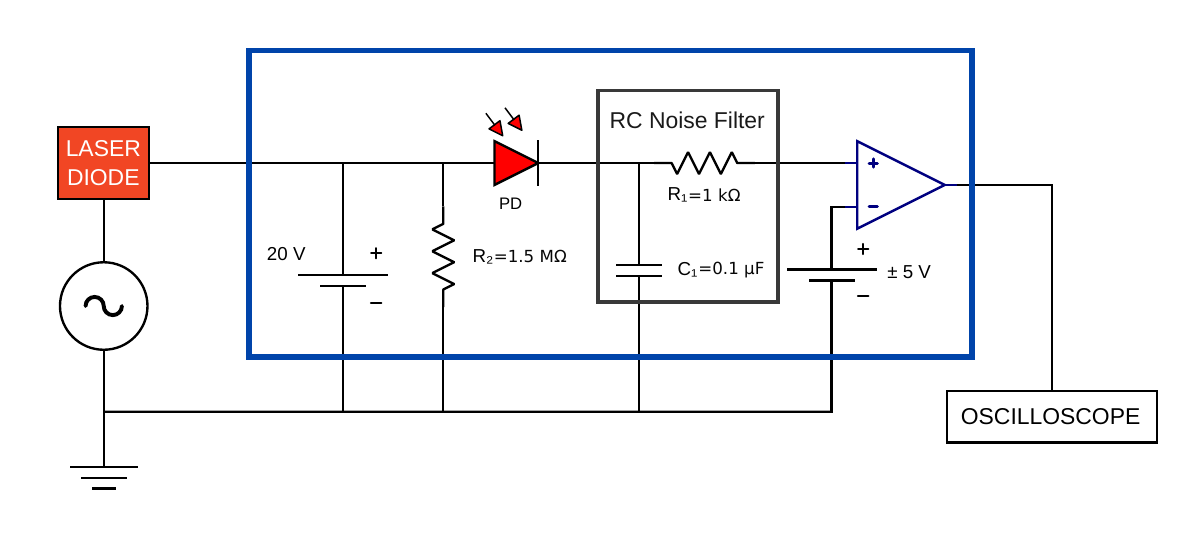}
    \caption{Schematics of the system used to characterize the diode. Starting from the left, a function generator powers the laser diode whose output is detected by a photodiode, coupled to an RC filter and an amplifier. The resulting signal is visualized with an oscilloscope.}
    \label{fig:rcamp_schematic}
\end{figure}

An LSP-1550-FC laser diode by Thorlabs was used to generate photons with an energy of $0.8$ eV.
An intermittent power supply was provided via a direct connection to a function generator to generate short laser pulses. To verify the proper operation of the device, the output of the laser was visualized with an oscilloscope, using a circuit made of a photodiode, an amplifier, and an RC filter as shown in \cref{fig:rcamp_schematic}.

To establish a connection between the diode, operating at room temperature, and the MKIDs located within the cryostat, an optical fiber with a $-20$ dB attenuator was employed.
For the optical fiber segment inside the cryostat, precautions were taken to prevent degassing and vacuum leakage. The PVC and nylon layers of the optical fiber were removed, exposing its coating and cladding. The end of the fiber was placed perpendicularly to the MKIDs chip at a distance of approximately 0.5 cm through a hole in the top of the chip holder. When operating the system, the short laser pulses, the attenuation and the geometrical acceptance of the resonators all contribute to exposing the detectors to a low mean amount of photons, which is necessary for the MKIDs to work in the photon-counting regime.

\begin{figure}
    \centering
    \includegraphics[width=0.95\textwidth]{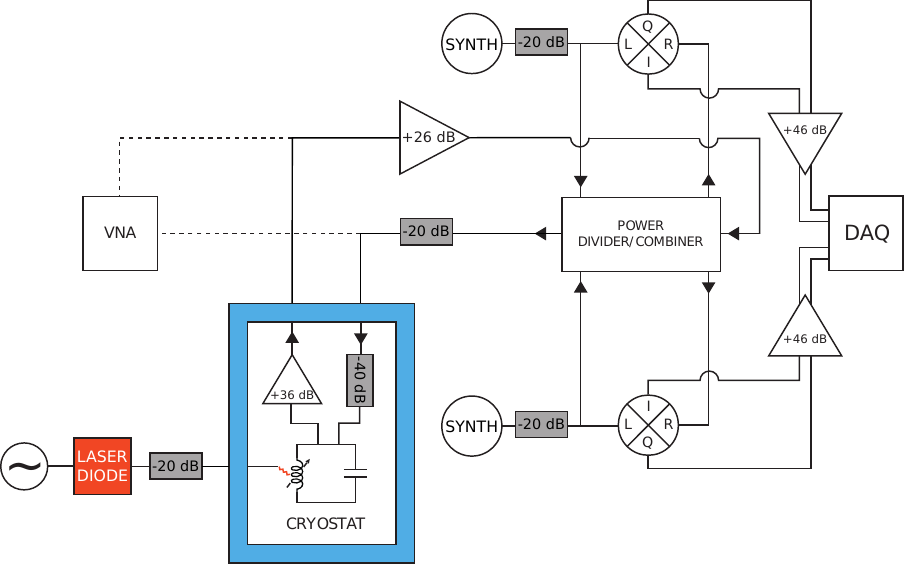}
    \caption{Circuit diagram for MKID characterization (dashed lines) and photon detection (solid lines). The latter allows for two resonators to be measured at the same time. The signal produced by a synthesizer is split and sent both to the local oscillator (LO) port of an IQ mixer and to the cryostat. The output from the cryostat is then sent to the IQ mixer, this time as RF, resulting in I and Q signals which are acquired by the data acquisition system (DAQ).}
    \label{fig:circuitsetup}
\end{figure}

In this experiment, the MKIDs response is probed through the heterodyne readout scheme shown in \cref{fig:circuitsetup}.
That is a type of mixer circuit designed to perform both amplitude and phase demodulation simultaneously, allowing for the extraction of both the in-phase (I) and quadrature (Q) components of a modulated signal.
In a heterodyne detection system, the same signal of frequency $f$ is sent through the circuit and the local oscillator (L) port of the mixer, while the transmitted signal is fed to the R, standing for RF, terminal. If the mixer is correctly calibrated, the resulting I and Q signals are proportional to the real and imaginary components of $S_{21}$.

The experimental procedure involves two distinct phases: detector characterization and photon detection. While the first one only required the use of a vector network analyzer (VNA) connected to the cryostat, the latter employed a custom circuit, shown in the scheme with solid lines.
In both cases, to adapt the room-temperature thermal noise to the cryogenic environment, the probe signal is attenuated by $-20$ dB at the $4$ K plate within the cryostat, and $-20$ dB at the mixing chamber, right before the MKIDs chip.
At the $4$ K stage, a high-electron-mobility transistor (HEMT) is employed to amplify the signal from the detectors by $+36$ dB, allowing for the acquisition of a measurable signal.

The tool used to characterize the MKIDs is an HP-8753 VNA, which directly measured the S-matrix components. On the other hand, the photon detection scheme is more complex and involves multiple instruments to handle rapid changes in the resonant frequency. Additionally, the custom circuit allows for the simultaneous readout of two distinct MKIDs, showcasing the scalability of the system.

Two FSW-0010 synthesizers by QuickSyn are employed to generate the probe signals, one for each resonator. The two signals are then combined and transmitted via a single microwave line to the cryostat.
The initial signal of $+15$ dB undergoes a total attenuation of $-80$ dB, split between $-40$ dB within the cryostat and $-40$ dB from external attenuation.
Additional amplification on top of that provided by the HEMT is given by a $+26$ dB RF amplifier placed outside of the cryostat.
The channels are then split, and the signals are down-converted with two IQ mixers using the synthesizer's pulses as Local Oscillator. Down-conversion is necessary as the data acquisition (DAQ) electronics used for the experiment (PXIe-5170R by National Instruments) operate in the MHz regime. Finally, the IQ components are further amplified by base-band amplifiers with a gain of $+46$ dB. Of the eight MKIDs present on the chip, only three have frequencies accessible with the employed VNA and heterodyne setup.

Although there are available software solutions~\cite{qcodes}, the students developed Python drivers for all the electronic instruments. To do so, they utilized common instrument libraries like PyVISA and PySerial~\cite{pyvisa,pyserial}. For the PXIe-5170R, they used the proprietary library Niscope~\cite{niscope}.

The decision to have students write the software for the data acquisition and analysis themselves was motivated by the desire to encourage learning in coding skills alongside their experimental work. 
\section{Experimental activities}\label{sec:activities}

\subsection{MKIDs characterization}\label{ssec:mkidchar}

The first part of the experiment focuses on studying the MKIDs resonances, measuring resonant frequencies and quality factors to estimate the superconductor material gap $\Delta$. To characterize the detector, this activity requires developing a robust fit procedure for the resonances profile, which will be essential for the correction of the raw data acquired in the latter photon-detection part of the experiment. Moreover, this activity can be taken as an example for characterizations of different kinds of superconducting resonators and could be proposed as a more general and self-contained laboratory experience if a custom MKID chip is not available.

The first step involves performing a raw estimate of the position of the resonances using the VNA while keeping the minimum temperature achievable by the setup, in our case $\sim 40$ mK. It is then possible to acquire their profiles as a function of the probe frequency and for different temperatures. 

Both the amplitude and phase of $S_{21}$ can be fitted using the module or the argument of \cref{eq:s21_for_fit} respectively. This equation is a variant of \cref{eq:s21_ideal} that includes additional parameters to account for non-ideal responses. These consist of a phase $\phi_0$ arising from mismatched impedances at the ports, which is responsible for the asymmetry of the resonance, and an overall complex constant $a$, which can be replaced by a low order polynomial to consider a non-flat background due to the entire readout line. 
\begin{figure}
	\begin{minipage}{0.52\linewidth}
		\centering
		\includegraphics[width=\linewidth]{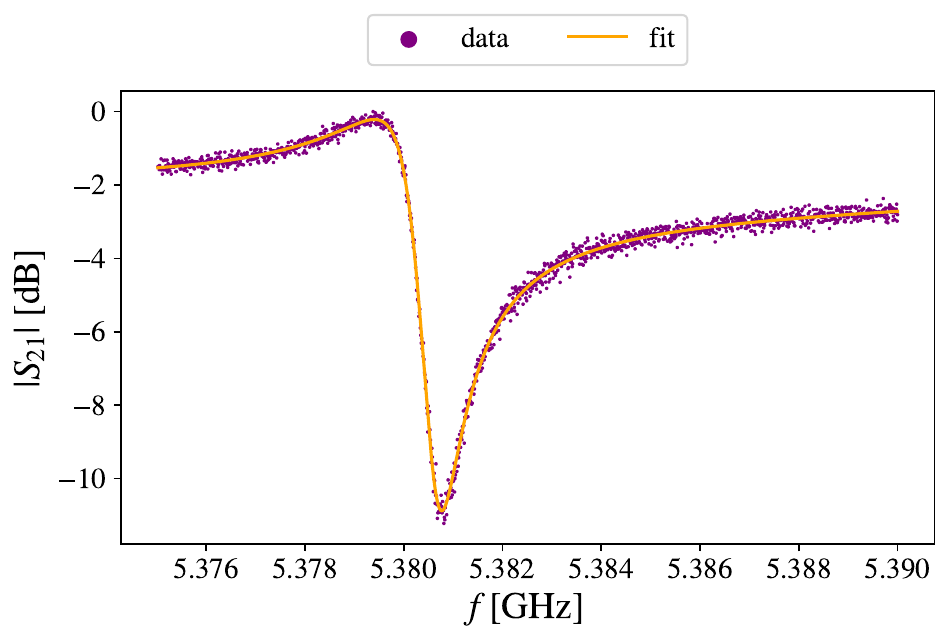}
	\end{minipage}\hfill
	\begin{minipage}{0.35\linewidth}
        \begin{tabular}{crr}
            \toprule
            Parameter & Value & Error\\
            \midrule
            $Q_{tot}$ & $4050$  & $20$\\
            $Q_i$ & $6440$  & $40$\\
            $Q_c$ & $10600$ & $140$ \\
            $\phi$ [rad]&$1.004$&$0.004$\\
            $f_0$ [MHz]&$5380.6$ & $0.1$\\
            \bottomrule
        \end{tabular}
	\end{minipage}
    \caption{Plot of one of the resonances, fitted with \cref{eq:s21_for_fit}. Relevant fitted parameters are shown in the table on the right.}
\end{figure}
\begin{equation}\label{eq:s21_for_fit}
    S_{21} = a\left( 1 - \frac{Q_{tot}}{Q_c} \frac{e^{j\phi_0}}{1+2jQ_{tot}\frac{f-f_0}{f_0}}\right)
\end{equation}
All the fits described in this work were performed with \texttt{iminuit}~\cite{iminuit}, which is a Python implementation of the \texttt{Minuit2} C++ library~\cite{minuit}.

To estimate $\Delta$, resonances were acquired at different temperatures ranging from $40$ mK up to $300$ mK in steps of $10$ mK. $Q_{tot}$ and $Q_c$ values were extracted from each fit, allowing for the determination of $Q_i$ using \cref{eq:q_relations}. The value of $\Delta$ was obtained using \cref{eq:invqi_for_gapfit}, where $\alpha$ was determined from a simulation of the circuit with \emph{Sonnet}~\cite{sonnet_manual}.

\begin{figure}
    \centering
    \subfloat[\centering \texttt{}]{{\includegraphics[width=0.55\textwidth]{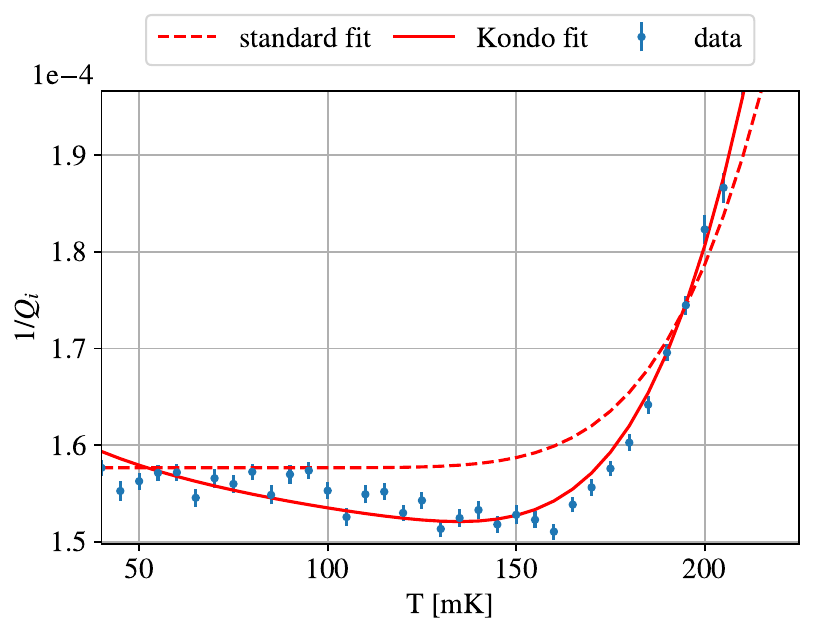}}}\hfill
    \subfloat[\centering\texttt{}]{{\includegraphics[width=0.41\textwidth]{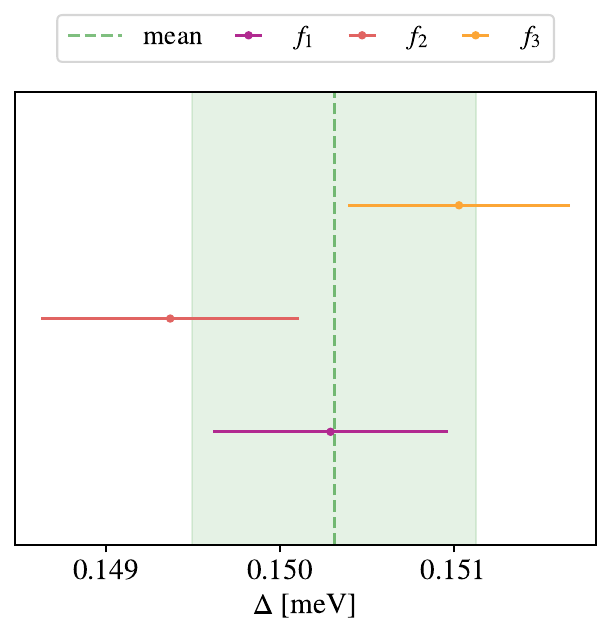}}}
    \caption{(a) Inverse of the internal quality factor as a function of temperature. The shift caused by the Kondo effect is clearly visible. (b) Fit results of the quality factors of three resonators and their weighted mean.}
    \label{fig:finalmeans}
\end{figure}

As apparent in the trend of the curve in \cref{fig:finalmeans}, the fit function does not perfectly describe the data points, suggesting that the model is incomplete. This trend can be attributed to the Kondo effect~\cite{Abrikosov1965,Hewson2009}, which is caused by the interaction of magnetic impurities with conduction electrons. This effect is taken into account by adding a logarithmic term to the fit model:

\begin{equation}\label{eq:kondo}
    \frac{1}{Q_i} = \frac{1}{Q_i(0)}+\frac{\alpha \sigma_1(T,\Delta)}{2\sigma_2(T, \Delta)} - b \ln \left( \frac{T}{T_K}\right)
\end{equation}
where the two newly introduced free parameters are a multiplication coefficient $b$ and the Kondo temperature $T_K$.

The results for $\Delta$ and $T_c$, shown in \cref{fig:finalmeans}, were obtained by considering the weighted mean of the values from three resonators, as the resonances of the other five MKIDs were outside the range of the VNA and thus could not be characterized:
\begin{eqnarray*}
        \Delta &= 0.150\pm 0.001\,\mathrm{meV} \\
        T_{c} &= 0.997 \pm 0.005\, \mathrm{K}
\end{eqnarray*}

\subsection{Development of the data acquisition system}

Another experimental activity that was carried out alongside the MKIDs characterization involved developing the DAQ system for the custom photon detection setup.

To generate the probe signals, the two FSW-0010 synthesizers were set to the resonant frequencies of the MKIDs with the highest $Q_{tot}$.

The PXIe-5170R digitizer by National Instruments, which can reach sampling frequencies up to $250$ MHz per channel, was used for data acquisition, employing all four channels ($I_1$, $Q_1$, $I_2$, $Q_2$) to exploit the multiplexing system. An analysis of the signals was performed to determine the acquisition parameters such as the trigger, number of records to store, record length, and sample rate.
Typical values used for data acquisition are $50$ MHz for the sampling frequency, $6000$ points for record length, and $1000$ records per acquisition. 
The digitizer ADCs (Analog to Digital Converters) were configured to work within the range of $\pm 5$ V and amplitude resolution of $14$ bits.

\begin{figure}
    \centering
    \subfloat[\centering \texttt{}]{{\includegraphics[width=0.5\textwidth]{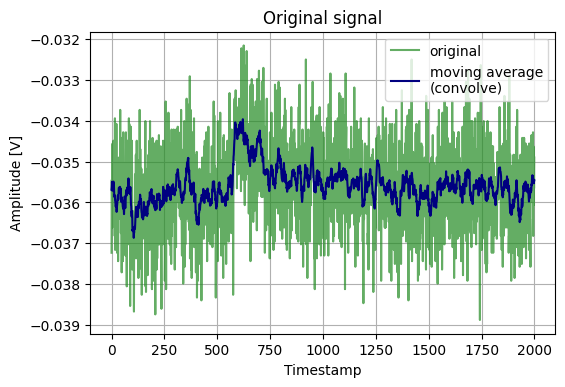}}\label{fig:mv_avg}}\hfill
    \subfloat[\centering\texttt{}]{{\includegraphics[width=0.47\textwidth]{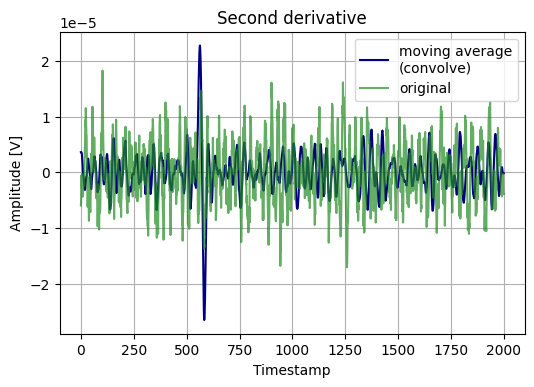}}    \label{fig:second_der}}
    \caption{Smoothing techniques applied to raw signals. (a) The green line shows a raw signal profile, while the blue one shows the moving averaged signal. (b) The second derivative of both the original and processed signal with the application of the Savitzky-Golay filter.}
\end{figure}

One of the most critical aspects of data acquisition is the trigger, which should efficiently select events while rejecting background noise or irrelevant processes. 
The most significant issue for the trigger used in this experiment was time jitter, which results in signals with the same amplitude triggering acquisition at different times.
To correctly align the signals, which is crucial for the following data analysis, a moving average smoothing technique was first applied to reduce noise, as shown in \cref{fig:mv_avg}. 

The Savitzky-Golay filter~\cite{SavGol} was then employed to smooth the second derivative of the signal.
The filter works by fitting a polynomial function to a moving window of adjacent data points and substituting the central point of the window with the fitted value. It is then possible to compute the signal n-th derivative from the polynomial fit.

The second derivative of double exponential-like signals consists of a fast, bipolar signal with a baseline of zero. This spike can be used to establish the starting point of each signal (\cref{fig:second_der}) and align them, reducing the time jitter.

\subsection{IQ correction}

The acquired raw data need to be corrected for the non-ideality of the experimental setup.
This includes different procedures to calibrate warm electronics, namely IQ mixers and cable delays, and to standardize the MKIDs response.

\begin{figure}
    \centering
    \subfloat[\centering \texttt{}]{{\includegraphics[scale=0.4]{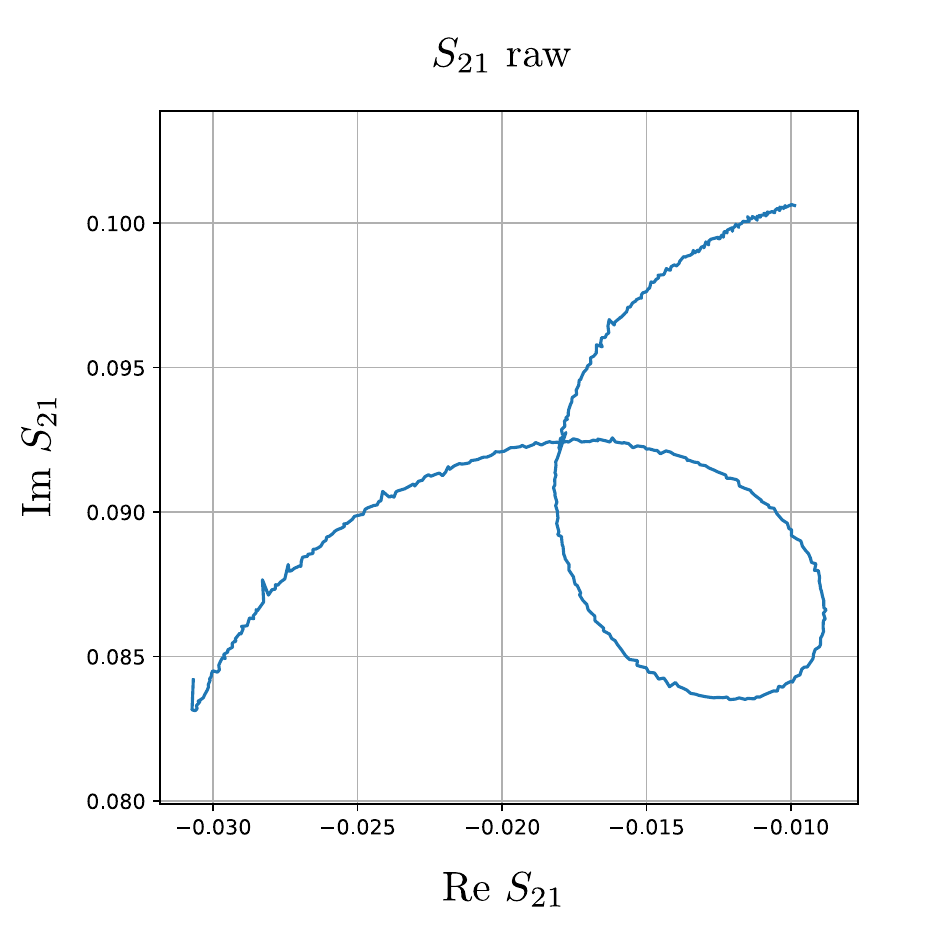}}}
    \subfloat[\centering\texttt{}]{{\includegraphics[scale=0.4]{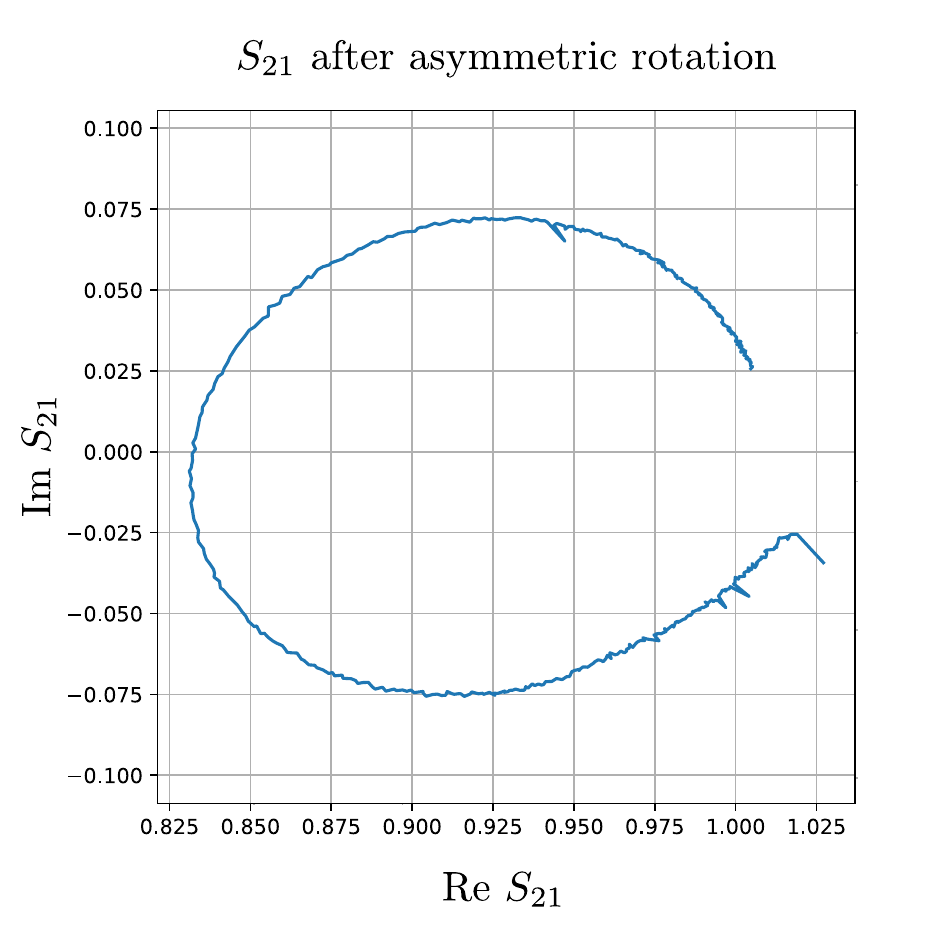}}}
    \caption{Correction of the resonance profile in the IQ plane. (a) Raw signal before applying IQ correction. (b) Corrected signal after applying IQ correction.}
    \label{fig:iqcorrection}
\end{figure}

The goal of this activity is to write the Python code required for transforming the raw acquired MKIDs resonances to correctly rotated and centered circles in the IQ plane. This correction will then be applied to each signal observed by the photon detectors. The following steps are employed:

\begin{description}
    \item[Cable delay effect correction:] both synthesizers and mixers are subject to a frequency-dependent offset.
    To address this issue, the RF port is disconnected, and the acquisition is performed for different LO frequencies.
    The resulting profile is subtracted from the signal of interest.
    
    \item[IQ mixer calibration:] Given sinusoidal Local Oscillator (LO) and RF signals with similar frequencies, the response of an ideal IQ mixer consists of a circular pattern in the IQ plane. In a realistic IQ mixer, however, the phase between I and Q is not precisely $90^\circ$ and the output is not balanced between the two ports. This results in observing an ellipse in the IQ plane. 
    To calibrate each IQ mixer, two synthesizers at slightly different frequencies are connected to the LO and RF ports. The resulting signal is fitted with an ellipse, which is then mapped to a circle.

    \item[Background correction:] even far from the MKIDs resonances, signals produced by the synthesizer arrive at the RF port with different amplitudes and phases. This leads to a frequency-dependent background which can be removed by fitting a wide-frequency scan around the resonance with the model of \cref{eq:s21_for_fit}, approximating the background as a polynomial. The fitted amplitude $A(\omega)$ and phase $\phi(\omega)$ are then used to correct the measured $S_{21}$ for the on-resonance frequencies:
        
    \begin{equation*}
        S_{21}(\omega) \rightarrow S_{21}(\omega)\frac{e^{-j\phi(\omega)}}{A(\omega)}
    \end{equation*}  
    
    \item[Center rotation]: each resonance is fitted with a circle, which is then rotated around its center to align the resonance's gap with the I-axis.
        
    \item[Asymmetry rotation:] the asymmetry in the shape of the resonance appears as a rotation of its circle in the IQ plane~\cite{khalil_analysis_2012}. To correct this effect, a counter-rotation and a contraction must be applied:
    \begin{equation*}
        S_{21} \rightarrow 1- \cos(\theta) e^{j\theta}(1-S_{21})
    \end{equation*}
\end{description}
The overall effect of these corrections is shown in \cref{fig:iqcorrection}.

\subsection{Photon-counting analysis}\label{sec:dataanalysis}

At this point, the data used for photon detection should consist of a series of pulses recorded by the trigger and corrected as detailed in the previous section. For each pulse, amplitude and phase are obtained from the I and Q signal. The analysis focuses on the observed phase pulses, whose height can be linked to the photon energy.

A more precise energy estimator than the simple peak-to-peak height can be obtained by applying the Optimum Filter (OF) method~\cite{Gatti1986}. The OF is able to achieve the best signal-to-noise ratio under the assumption that the signal is a linear combination of a mean signal and ergodic noise. This method allows the construction of a frequency filter that suppresses the frequencies where the noise contribution is more significant.
There are various implementations of the filter, with the one used in this case being based on the Discrete Fourier Transform (DFT). 

\begin{figure}
    \centering
        \subfloat[\centering \texttt{}]{{\includegraphics[width=0.43\textwidth]{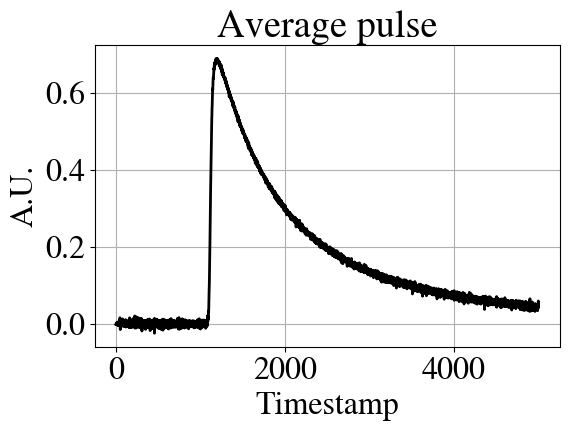}}}\hfill
        \subfloat[\centering\texttt{}]{{\includegraphics[width=0.43\textwidth]{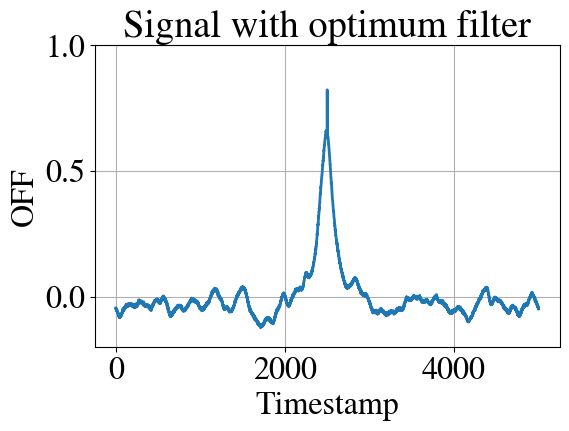}}}
        \caption{Example of Optimum Filter application. (a) Average pulse obtained from $2000$ signals. (b) A signal filtered with the OF. The result is a cusp, the function with the best signal-to-noise ratio.}
        \label{fig:OF_results}  
    \centering
\end{figure}

The OF takes in input some samples of noise and signals and outputs the frequency spectrum of the noise, the average pulse (average of the signals, \cref{fig:OF_results} (a)), and a transfer function that can be applied to the signals to reduce their noise component.
The transfer function is then applied to the signals and the mean of the DFT of the filtered signal (OFF) is used as an estimator of the measured energy.
An example of a signal after this step is shown in \cref{fig:OF_results} (b).

\begin{figure}
    \centering
    \includegraphics[width=0.6\textwidth]{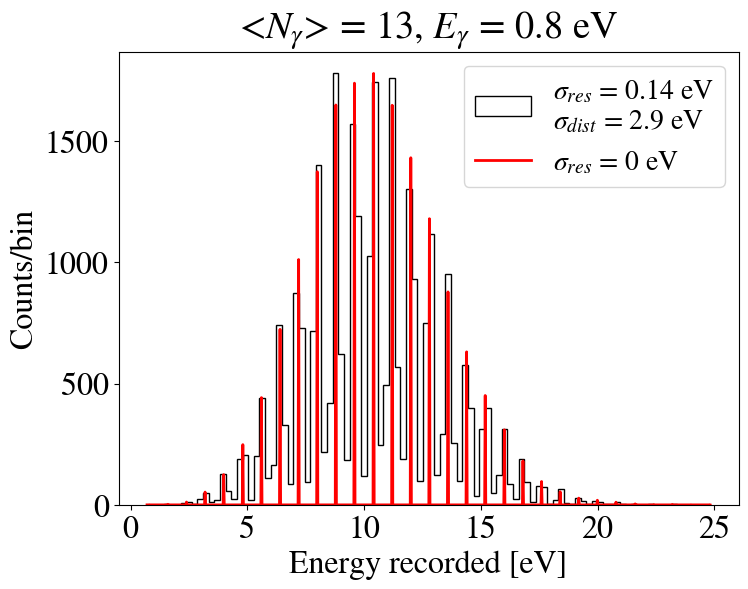}
    \caption{Simulation of the expected distribution of the amplitudes in the ideal case (red) with an infinite detector resolution and in a real case (black) with a finite resolution good enough to differentiate among the different peaks. $\sigma_\mathit{res}$ is the detector resolution and $\sigma_\mathit{dist}$ is the standard deviation of all the samples. In both cases, different peaks are visible and distinguishable.}
    \label{fig:simulation}
\end{figure}

The expected energy spectrum is shown in \cref{fig:simulation}.
The laser light reaching the detector consists of a coherent state given by a superposition of different photon-number states, which are distributed according to Poisson statistics with average $\mu$. Since each event observed by the detector arises from a definite number of photons being absorbed by the resonator, the spectrum reconstructed by an ideal detector with infinite energy resolution reflects the structure of the coherent state, consisting of a series of discrete peaks at integer multiples of the laser's single-photon energy.

Assuming a Gaussian energy spread due to the finite detector's resolution, the observed spectrum is given by a sum of Gaussians weighted by a Poissonian distribution. 

The expected spectrum can be used as a fit function for the observed data, which are binned for different values of the OFF:
\begin{equation}\label{eq:Gaussfit}
    \textrm{model}(\textrm{OFF}, \sigma, \mu, A, \delta, E_{\gamma}) = A \sum_{n=0}^{N_{max}} P(n, \mu) \cdot G(\textrm{OFF}, nE_{\gamma} + \delta, \sigma)
\end{equation}
where $A$ is a normalization factor, $N_{max}$ is a fixed maximum number of considered photons, $\delta$ is a global shift of the fit function from zero, $G$ is the Gaussian distribution, $P$ is the Poisson distribution and $E_{\gamma}$ is the conversion from A.U. to eV that can be obtained knowing the energy of one infrared photon, i.e. $0.8$ eV.

\begin{figure}
    \begin{minipage}
    {0.5\linewidth}
    \flushleft\includegraphics[width=\linewidth]{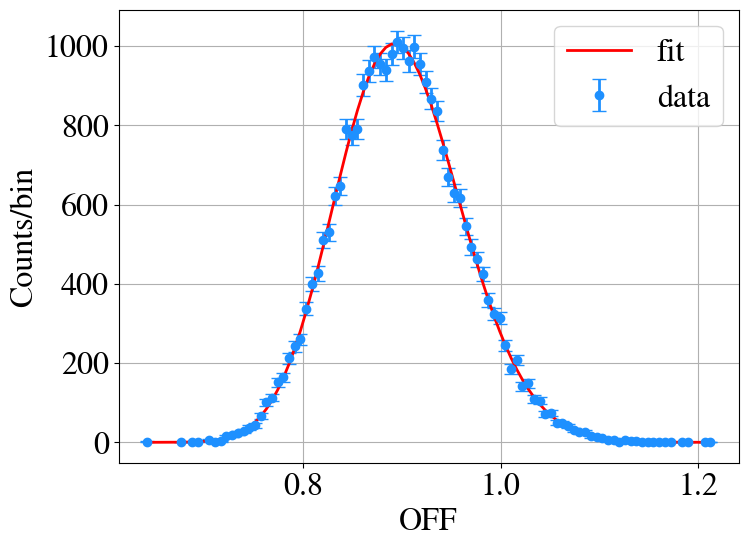}
	\end{minipage}
    \begin{minipage}{0.35\linewidth}
    \begin{tabular}{c r r r}
        \toprule
        Parameter   &   Value	    &    Error	  &   Constraint    \\
        \midrule
        $\sigma$     &	27.6e-3 	&    -        &	  fixed           \\
        $\mu$     	&   12.4	    &    1.6	  &   $>0$            \\
        $A$     	&   159.2	    &    0.9	  &	  $> 10^{-5}$	  \\
        $\delta$     &	0.700	    &    0.013	  &	  $\in [0, 1]$    \\
        $E_\gamma$  &  	0.016	    &    0.001    &   -               \\
        \bottomrule
    \end{tabular}
	   \end{minipage}\hfill
    \caption{Histogram of the amplitudes computed with the OF. Data points are in blue, with error bars obtained assuming Poisson fluctuations in each bin, and the fit function is in red. The table on the right shows the best parameters found with the fit, with $\chi^2/$dof $=1.2$. The last column shows the parameters' constraints.}
    \label{fig:result}
\end{figure}

The measured distribution is shown in \cref{fig:result}. The peaks due to different photon numbers cannot be separated, likely due to the detector resolution being much larger than the photon energy.

The average number of photons can still be estimated with the fit.
Due to the large number of parameters, it can be difficult to make the fit converge and more than one combination of parameters can give good $\chi^2$ values.
To avoid that, some parameters can be constrained. For example, $\sigma$ is set to the detector's resolution, which is calculated as the Root Mean Squared (RMS) of the noise signals (after the OF) since it is the dominant component.
All constraints are shown in the table in \cref{fig:result}.
The result of the fit, with a $\chi^2/$dof$=1.2$, is shown in \cref{fig:result}.
The best estimate for the average number of photons is $12 \pm 2$. 

\section{Conclusions}

This paper presented a laboratory experiment designed for graduate students in the field of quantum technologies, which aims to measure the number-resolved energy spectrum of photons in the far-infrared range using Microwave Kinetic Inductance Detectors (MKID).
The experiment involves characterizing the resonators and developing a data acquisition system to acquire the signals, concluding with the data analysis of the measurements using the characterized MKIDs. 

The experience was first realized at the University of Milano-Bicocca in 2022 and the students reported an average number of observed photons of $12\pm2$, even if the peaks due to different numbers of photons could not be resolved.
Improvements in the experimental setup and the fabrication process of the MKIDs could enhance the resolution, allowing the distinction of different numbers of interacting photons as separate peaks in the energy spectrum.

Given the students' positive feedback regarding the course, particularly highlighting the exposure to various new topics such as cryogenics, light-matter interaction, and superconductivity, as well as the acquisition of skills in coding, data analysis, and problem-solving, the experiment has evolved into the primary student-led initiative within the new laboratory course at the cryogenics laboratories of the University of Milano-Bicocca. This trend has persisted throughout the years 2023 and 2024 and is anticipated to continue.

\section*{Acknowledgments}
Special thanks go to our laboratory supervisors for the opportunity and their guidance in our experimental activities. We acknowledge Benno Margesin and the FBK clean-room team for the chip fabrication. Additionally, we also acknowledge Renato Mezzena and Andrea Vinante for providing the device. 
\section*{Data availability}
The data that support the findings of this study are available upon request from the authors.
\section*{Disclaimer}
Mention of commercial products is for information only and does not imply a recommendation or endorsement.
\printbibliography
\end{document}